# QED DERIVED FROM THE TWO-BODY INTERACTION (1)

*MAPPING TO M-SPACE*

Sarah B. M. Bell,[1,2] John P. Cullerne,[1] Bernard M. Diaz[1,3]

## Abstract

We have shown in a previous paper that the Dirac bispinor can vary like a four-vector and that Quantum Electrodynamics (QED) can be reproduced with this form of behaviour.[(1)] Here, in part (1) of this paper, we show that QED with the same transformational behaviour also holds in an alternative space we call *M*-space. We use the four-vector behaviour to model the two-body interaction in *M* and show that this has similar physical properties to the usual model in *L* which it predicts. In part (2) of this paper[(2)] we use *M*-space to show that QED can be reduced to two simple rules for a two-body interaction.

## 1. INTRODUCTION

### 1.1 Note on nomenclature

‡ signifies quaternion conjugation. A lowercase Latin subscript stands for 1, 2 or 3 and indicates the space axes. A lowercase Greek subscript stands for 0,

---

[1] Department of Computer Science I.Q. Group, The University of Liverpool, Chadwick Building, Peach Street, Liverpool, L69 7ZF, United Kingdom.

[2] U.K. phone number and email address, 01865 798579 and Sarabell@dial.pipex.com

[3] E-mail, B.M.Diaz@csc.liv.ac.uk

1, 2, or 3 and indicates the spacetime axes. $i = \sqrt{-1}$. $\mathbf{i}_0 = 1$. $\mathbf{i}_1 = \mathbf{i}$, $\mathbf{i}_2 = \mathbf{j}$, $\mathbf{i}_3 = \mathbf{k}$ stand for the quaternion matrices, where $\mathbf{i}_r^2 = -1$, $\mathbf{i}_1\mathbf{i}_2 = \mathbf{i}_3$, $\mathbf{i}_2\mathbf{i}_1 = -\mathbf{i}_3$ with cyclic variations. $\mathbf{i}_r^\ddagger = -\mathbf{i}_r$. $\mathbf{i}_0^\ddagger = \mathbf{i}_0$. Our base for the quaternion matrices is $\mathbf{i}_r = -i\boldsymbol{\sigma}_r$ where $\boldsymbol{\sigma}_r$ are the Pauli matrices. We set the speed of light, c, and $h/2\pi$, where $h$ is Planck's constant, to 1 except for final results.

## 2.1 Summary

We have shown in a previous paper that the Dirac equation may be solved to yield a wave function that is a reflector matrix whose quaternion elements are the Dirac bispinors.[1] In this novel representation of a bispinor, it is easy to show that under spatial rotation, Lorentz transformation, charge conjugation, parity change and time reversal, both the bispinor and the mass term can vary like a four-vector. There is also a conserved charge density current where these variables have the same behaviour, which leads to a photon equation. The Dirac and photon equations hold in the usual four-dimensional flat spacetime which we call *L*. Here we will show that the Dirac and photon equations with the same transformational behaviour also hold in a second four-dimensional space we call *M*. *M* has a scalar parameter, *R*, describing the space and an alternative topology.

We define a bijection between points in *L* and points in *M* which allow us to determine the behaviour of a particle in *M* given its behaviour in *L* or the reverse. Using this, we discover how to relate the charge density current in *M* to that in *L* which means we can transfer an interaction in *L* into *M* or the reverse.

We describe the two-body attractive interaction by assuming one body generates a potential field in conformity with the photon equation while the behaviour of the second body is determined by solving the Dirac equation with this potential. We discover how to model the attractive two-body interaction in *M* using the photon and Dirac equations. We show that the solutions of the Dirac equation resemble the orbits of Bohr.[6, 7] We deduce the same two equations as Bohr did to describe such two-body interactions although we transfer them into the new space *M*. We call these equations *the Bohr equations*. We call our model of the two-body interaction in *M* the Bohr interaction.

We resolve the discrepancy between the integer angular momentum for the two-body interaction predicted by the Bohr equations for *M* and the half-integer angular momentum found when the Dirac equation is solved in the usual way in *L*, for example the one-electron atom.[5] We will call this latter structure *the Dirac interaction*. We show that the two methods of describing the two-body interaction lead to similar results, and that in many circumstances the two states would be indistinguishable.

In part (2) of this paper[2] we use the results here to derive QED from the two-body interaction. We use a space whose local topology and curvature varies. However, this space is locally equivalent to some *M*, and continues to map to *L*.

## 2. THE EQUATIONS IN *M*-SPACE

### 2.1 Preliminaries

The versatile Dirac equation for an electron is[1]

$$(\underline{\mathbf{D}} - ie\underline{\mathbf{A}}^\sim)\underline{\Phi} = \underline{\Phi}\underline{\mathbf{M}} \qquad (2.1.\text{A})$$

where *e* is the charge on the electron, and we define the meaning of the underline by

$$\underline{\mathbf{D}} = \underline{\mathbf{D}}(\mathbf{D}, \mathbf{D}^\ddagger) = \begin{pmatrix} 0 & \mathbf{D} \\ \mathbf{D}^\ddagger & 0 \end{pmatrix} \qquad (2.1.\text{B})$$

We call such matrices *reflectors*. We also have

$$\mathbf{D}_0 = i\mathbf{i}_0 \, \partial/\partial x_0, \qquad \mathbf{D}_r = \mathbf{i}_r \, \partial/\partial x_r, \qquad (2.1.\text{C})$$

where the $x_\mu$ are real. Then

$$\mathbf{D} = \mathbf{D}_0 + \mathbf{D}_1 + \mathbf{D}_2 + \mathbf{D}_3 \qquad (2.1.\text{D})$$

The Dirac equation (2.1.A) holds in *L*, which we have co-ordinated with the Cartesian co-ordinates $(x_0, x_1, x_2, x_3)$. We provide a definition for $\underline{\mathbf{A}}^\sim$, the potential term, similar to that for $\underline{\mathbf{D}}$

$$\underline{\mathbf{A}}^\sim = \underline{\mathbf{A}}^\sim(\mathbf{A}^\sim, \mathbf{A}^{\sim\ddagger}), \qquad \mathbf{A}^\sim = A_0/i + \mathbf{i}_1 A_1 + \mathbf{i}_2 A_2 + \mathbf{i}_3 A_3 \qquad (2.1.\text{E})$$

where the four-vector $(A_0, A_1, A_2, A_3)$ represents the potential, while

$$\underline{\Phi} = \underline{\Phi}(\phi_1, \phi_2) \qquad (2.1.\text{F})$$

where $\underline{\Phi}$ is the wave function, $\phi_1$ and $\phi_2$ are quaternions whose relation to the bispinors in the original version of the Dirac equation is described by Bell et al.,[1] and

$$\underline{\mathbf{M}} = \underline{\mathbf{M}}(\mathbf{M}, -\mathbf{M}^{\ddagger}) \tag{2.1.G}$$

$$\mathbf{M} = \mathbf{i}_0 M_{\tilde{0}} + \mathbf{i}_1 M_1 + \mathbf{i}_2 M_2 + \mathbf{i}_3 M_3, \quad \mathbf{MM}^{\ddagger} = -m_e^2$$

where $M_{\tilde{0}}$ is imaginary and $m_e$ is the rest mass of the electron. For the usual half-angular behaviour of the phase of the bispinor, $\phi_1$, under rotation and the analogue under Lorentz transformation, $\mathbf{M} = -im_e$, a constant. $\phi_2$ also shows half-angular behaviour, but it rotates in the opposite sense to $\phi_1$ for the temporal rotations which implement Lorentz transformations for the versatile Dirac equation. This means that $\phi_2^{\ddagger}$ rotates in the same sense as $\phi_1$. In this case the versatile Dirac equation is simply another form of the usual version. However, the versatile Dirac equation also allows four-vector behaviour of the bispinors $\phi_1$ and $\phi_2^{\ddagger}$. In this case $(iM_{\tilde{0}}, M_1, M_2, M_3)$ behaves like the energy-momentum four-vector of the electron.

## 2.2. Definition of *M*-space

We transform the photon and Dirac equations into the alternative space, *M*. We define *M*-space in terms of an explicit co-ordination of *M* and *L* which highlights the simple symmetry we want to retain.

The $(x_1, x_2)$ plane of *L* may also be described using polar co-ordinates, $(r, \theta)$, where we have

$$x_1 = r\sin\theta, \qquad x_2 = r\cos\theta \tag{2.2.A}$$

$$s' = r\theta, \qquad \overrightarrow{\delta s'} = r\overrightarrow{\delta\theta} \tag{2.2.B}$$

and $s'$ is the arc corresponding to $r$ and $\theta$ in $L$. We will also use the alternative set of co-ordinates for $L$, $(x_0, s', r, x_3)$. Proceeding in the standard way we have

$$\frac{\partial}{\partial x_1} = \sin\theta \frac{\partial}{\partial r} + \frac{1}{r}\cos\theta \frac{\partial}{\partial \theta}, \qquad (2.2.C)$$

$$\frac{\partial}{\partial x_2} = \cos\theta \frac{\partial}{\partial r} - \frac{1}{r}\sin\theta \frac{\partial}{\partial \theta}$$

We introduce a further vector along the arc

$$s = R\theta, \qquad \overrightarrow{\delta s} = R\overrightarrow{\delta \theta} \qquad (2.2.D)$$

where $R$ is constant. For reasons that will become apparent, we call $R$ *the Bohr radius*. Using the first of equations (2.2.D), equations (2.2.C) become

$$\frac{\partial}{\partial x_1} = \sin\theta \frac{\partial}{\partial r} + \frac{s}{s'}\cos\theta \frac{\partial}{\partial s}, \qquad (2.2.E)$$

$$\frac{\partial}{\partial x_2} = \cos\theta \frac{\partial}{\partial r} - \frac{s}{s'}\sin\theta \frac{\partial}{\partial s}$$

We transform the four-vectors associated with **A** and **M**, with

$$A_1 = A_r \sin\theta + A_s \cos\theta, \qquad A_2 = A_r \cos\theta - A_s \sin\theta \qquad (2.2.F)$$

$$M_1 = M_r \sin\theta + M_s \cos\theta, \quad M_2 = M_r \cos\theta - M_s \sin\theta$$

where $A_r$ and $M_r$ are the components along the radius, $\overrightarrow{\delta r}$, and $A_s$ and $M_s$ are the components along the arc, $\overrightarrow{\delta s}$. We transform the matrices, $\mathbf{i}_1$ and $\mathbf{i}_2$, with

$$\mathbf{i}_1 = \mathbf{i}_r \sin\theta + \mathbf{i}_s \cos\theta, \qquad (2.2.G)$$

$$\mathbf{i}_2 = \mathbf{i}_r \cos\theta - \mathbf{i}_s \sin\theta$$

where $\mathbf{i}_r$ and $\mathbf{i}_s$ are the quaternion matrices along the radius, $\vec{\delta r}$, and the arc, $\vec{\delta s}$, respectively. The multiplication table and commutation relations of the matrices $(\mathbf{i}_0, \mathbf{i}_s, \mathbf{i}_r, \mathbf{i}_3)$ are the same as that of the matrices $(\mathbf{i}_0, \mathbf{i}_1, \mathbf{i}_2, \mathbf{i}_3)$, respectively, as may be easily checked. From the second of equations (2.1.E), and (2.1.G), and equations (2.2.F) and (2.2.G) we obtain

$$\tilde{\mathbf{A}} = -i\mathbf{i}_0 A_0 + \mathbf{i}_s A_s + \mathbf{i}_r A_r + \mathbf{i}_3 A_3, \qquad (2.2.\mathrm{H})$$

$$\mathbf{M} = \mathbf{i}_0 \tilde{M_0} + \mathbf{i}_s M_s + \mathbf{i}_r M_r + \mathbf{i}_3 M_3$$

When $r = R$, $s' = s$ and we have from equations (2.1.C), (2.2.E) and (2.2.G)

$$\mathbf{D}_1 + \mathbf{D}_2 = \mathbf{i}_s \partial/\partial s + \mathbf{i}_r \partial/\partial r \qquad (2.2.\mathrm{I})$$

while $\mathbf{D}_0$ and $\mathbf{D}_3$ are unchanged. We would like equation (2.2.I) to hold universally. We may do this by altering equations (2.2.E) so that the first term still applies but the multiplier, $s/s'$, of the second term is removed. $\vec{\delta r}$ is orthogonal to both $\vec{\delta s'}$ and $\vec{\delta s}$, from the second of equations (2.2.B) and (2.2.D). We may define a new space, $M$, by retaining $r$ but setting $s'$ to $s$. The co-ordinates applicable to $M$ are then $(x_0, s, r, x_3)$. The first of equations (2.2.B) and (2.2.D) provide a bijection between $L$ and $M$ by relating the one co-ordinate not shared

$$s' = \frac{r}{R} s \qquad (2.2.\mathrm{J})$$

From equations (2.2.H) and (2.2.I) we can translate the Dirac equation into $M$ immediately.

We can visualise the space *M* by supposing that the length of an observer's measuring rod in *M* in the direction of an arc, $\vec{\delta s}$, is proportional to *r* while the length remains constant for radial distances. Equation (2.2.J) shows that *M* is well-behaved and we may calculate the metric tensor by considering small displacements, $\delta \tau_M$.[12] Using equation (2.2.J), we have for *L* and *M* respectively

$$d\tau_L^2 = dx_0^2 + r^2 d\theta^2 + dr^2 + dx_3^2, \qquad (2.2.K)$$

$$d\tau_M^2 = dx_0^2 + R^2 d\theta^2 + dr^2 + dx_3^2$$

where $d\tau_L$ is a small displacement in *L* and we have used the same co-ordinates for both spaces for the purposes of comparison. It is apparent by inspection of the metric for *M* that the space is not curved in the sense applicable to General Relativity.[12] Nevertheless, the mapping (2.2.J) gives it rich properties most usually associated with curvature.

Suppose we have an application for the Dirac equation (2.1.A) in *L*. We would like to know how to formulate the same application in *M*. We have translated the variables into the co-ordinate system for *M*, but we have not yet eradicated all the effects of the mapping (2.2.J). Since the same volume has a different size in *M* and *L* under the mapping, volumes will also be affected and so will any variables defined using volumes. **M** has no dependency on volume and is not affected but the potential is derived from a Dirac current. The latter is a charge density current and therefore both the current and the potential are affected if the volume element changes. The wave function is used to define another Dirac current and will also be affected.

We calculate the effect of the change of volume element. The volume

element for *L* is

$$\delta V^L = \delta x_1 \delta x_2 \delta x_3 \quad (2.2.\text{L})$$

We consider the volume element where $\overrightarrow{\delta x_1}$ lies along $\overrightarrow{\delta s'}$, $\overrightarrow{\delta x_2}$ lies along $\vec{r}$, $\delta x_1 = \delta s'$ and $\delta x_2 = \delta r$,

$$\delta V^L = \delta s' \delta r \delta x_3 \quad (2.2.\text{M})$$

From equations (2.2.B) and (2.2.D) the same volume element for *M* is

$$\delta V^M = \delta s \delta r \delta x_3 \quad (2.2.\text{N})$$

From equations (2.2.B), (2.2.D), (2.2.M) and (2.2.N) we see that

$$\delta V^M = \frac{R}{r} \delta V^L \quad (2.2.\text{O})$$

Since the volume element, $\delta V^L$, is of constant magnitude in *L* and the volume element, $\delta V^M$, is of constant magnitude in *M*, equation (2.2.O) holds for all cases.

### 2.3. Revising the potential for *M*-space

Proceeding as in Bell et al.[1] we find the versatile photon equation for *L*. It is

$$\underline{\mathbf{D}}\underline{\mathbf{D}}\tilde{\mathbf{A}} = \tilde{\mathbf{J}} \quad (2.3.\text{A})$$

where

$$\underline{\tilde{\mathbf{J}}} = \underline{\tilde{\mathbf{J}}}(\tilde{\mathbf{J}}, \tilde{\mathbf{J}}^{\ddagger}), \qquad \tilde{\mathbf{J}} = \tilde{J_0}\mathbf{i}_0 + \sum_r \tilde{J_r}\mathbf{i}_r, \quad (2.3.\text{B})$$

and $\mathbf{J}^L = (iJ_0^\sim, J_1, J_2, J_3)$ is the usual Dirac current.

The only property of interest here is that for a suitably chosen frame the current may be expressed as a charge, $P'(x_0, r, \theta, x_3)$, per unit volume element. Using a Lorentz transformation, $Z(x_0, r, \theta, x_3)$, we transform equation (2.3.A) to this frame. From equations (2.3.A) and (2.3.B) we obtain

$$\mathbf{DD}^\ddagger A = P'/\delta V^L \qquad (2.3.\text{C})$$

where the potential in $L$ is $A = iA^\sim$. Since $\delta V^L$ is constant in $L$

$$\mathbf{DD}^\ddagger A\, \delta V^L = P' \qquad (2.3.\text{D})$$

Equation (2.3.A) and (2.3.B) are replaced by their analogues for $M$ using the method we followed for the versatile Dirac equation. In particular we have

$$\mathbf{J}^\sim = \mathbf{i}_0 J_0^\sim + \mathbf{i}_s J_s + \mathbf{i}_r J_r + \mathbf{i}_3 J_3, \qquad (2.3.\text{E})$$

where $J_s$ and $J_r$ are the components along the arc and radius. $\delta V^L$ must be exchanged for $\delta V^M$ in equation (2.3.C) and, since $\delta V^M$ is constant for $M$, we obtain instead of equation (2.3.D)

$$\mathbf{DD}^\ddagger A^M\, \delta V^M = P' \qquad (2.3.\text{F})$$

where the potential in $M$ is $A^M = iA^{M\sim}$ and $P'$ must be the same for both $M$ and $L$ because it refers to the same point.

Setting the left-hand sides of equations (2.3.D) and (2.3.F) equal and using equation (2.2.O) we have in the general case

$$A^M = \frac{Ar}{R} \tag{2.3.G}$$

We transform $A^{M\sim}$ using $Z^{-1}$ to find $\mathbf{A}^{M\sim}$ and we quote the final form of the potential term for *M*

$$\mathbf{A}^{M\sim} = \mathbf{i}_0 A_0^{M\sim} + \mathbf{i}_s A_s^M + \mathbf{i}_r A_r^M + \mathbf{i}_3 A_3^M \tag{2.3.H}$$

We now know that, if $\mathbf{A}^{\sim}$ is the appropriate potential term for some application in *L*, then $\mathbf{A}^{M\sim}$ is the appropriate potential term in *M*.

The final version of the Dirac equation in *M* is

$$\{\underline{\mathbf{D}}(\mathbf{D},\mathbf{D}^{\ddagger}) - ie\underline{\mathbf{A}}^{M\sim}(\mathbf{A}^{M\sim},\mathbf{A}^{M\sim\ddagger})\}\underline{\boldsymbol{\Phi}}^M(\phi_1^M,\phi_2^M) \tag{2.3.I}$$
$$= \underline{\boldsymbol{\Phi}}^M(\phi_1^M,\phi_2^M)\underline{\mathbf{M}}(\mathbf{M},-\mathbf{M}^{\ddagger})$$

where $\underline{\boldsymbol{\Phi}}^M$ is the solution associated with the new potential term $\underline{\mathbf{A}}^{M\sim}$. We call the transformation of equation (2.1.A) for *L* into equation (2.3.I) for *M the Circular transformation*. The Circular transformation includes the exchanging co-ordinates $(x_1, x_2)$ for $(s, r)$ and any other adjustment to the variables required because of the mapping from *L* to *M*, in this case an adjustment to $\mathbf{A}^{\sim}$. The Dirac and photon equations have the same form and significance for an observer in *M* as for an observer in *L* after the map

$$s \to x_1, \qquad r \to x_2, \qquad \mathbf{A}^{M\sim} \to \mathbf{A}^{\sim} \tag{2.3.J}$$

We see that *M* differs topologically from that of *L*. Apart from this, the co-ordinates $(x_0, s, r, x_3)$ appear to an observer in *M* to be a Cartesian system in a flat spacetime. He therefore sees a flat metric with Cartesian co-ordinates

$$d\tau_M^2 = dx_0^2 + ds^2 + dr^2 + dx_3^2 \qquad (2.3.K)$$

We note in particular that, since the photon and Dirac equation hold in *M*, conservation of energy is guaranteed for electromagnetic processes.

## 3. TWO-BODY SOLUTION IN *M*

### 3.1 Bohr's first equation

We consider a two-body attractive interaction in *M*. We work in the rest frame of one body which we call *the nucleus*. We may solve the photon equation in its original form, equation (2.3.A), in the usual way for the potential due to the nucleus, $iA\tilde{}$, in *L*. This gives

$$A\tilde{} = \frac{if}{a} \qquad (3.1.A)$$

where *f* represents the charge on, and *a* the distance from, the nucleus. Restricting (3.1.A) to a plane co-ordinated by $(s, r)$ and assuming the nucleus is at the origin we obtain

$$A\tilde{} = \frac{if}{r} \qquad (3.1.B)$$

We suppose that all values of the $x_3$ co-ordinate for the nucleus are equally likely ab initio, but that if we fix the $x_3$ co-ordinate of the electron then we fix the $x_3$ co-ordinate of the nucleus as having the same value. In that case equation (3.1.B) holds in *L* for all $x_3$. We shall retrieve the usual model based on equation (3.1.A) in section 4.4. From equations (3.1.B) and (2.3.G)

$$A^{M\sim} = \frac{if}{R} \qquad (3.1.C)$$

where $iA^{M\sim}$ is the potential due to the nucleus in *M*.

To be concrete we assume the second body is an electron. Since the electron is in a constant potential, $iA^{M\sim}$, we expect the solution to the Dirac equation to be a plane wave. In the most general case the electron will have a velocity but for a plane wave this must be constant. We add the condition that if the electron is near the nucleus it must remain so, on the principle that we expect an attractive interaction to yield a bound state. This means that the electron cannot have any radial velocity and the velocity must lie along $\vec{\delta s}$, that is, the electron has a circular orbit round the nucleus. This leads to a solution of equation (2.3.I)

$$\phi_1^M = \exp\{i(v\tilde{\ }\tilde{x_0} + \mu s)\}, \qquad (3.1.D)$$
$$\phi_2^M = i\{v\tilde{\ } - \mathbf{i}_s\mu - eA^{M\sim}\}\mathbf{M}^{-1}\exp\{i(v\tilde{\ }\tilde{x_0} + \mu s)\},$$
$$m^{\sim 2} = (v\tilde{\ } - eA^{M\sim})^2 + \mu^2$$

where $m\tilde{\ } = m_e/i$, $\tilde{x_0} = x_0/i$, $\underline{\Phi}^M(\phi_1^M, \phi_2^M)$ is the wave function, $iv\tilde{\ }$ is the frequency and $\mu$ the wave number of the electron. This bound state for the two-body attractive interaction in *M* defines *the Bohr interaction*.

We see from the third of equations (3.1.D) that there is an equivalent free electron with wave number $\mu$ and frequency $i\eta\tilde{\ }$ given by

$$\eta\tilde{\ } = v\tilde{\ } - eA^{M\sim} \qquad (3.1.E)$$

This electron, which we call *the Bohr electron*, can be seen as the one originally considered by Bohr in his model for the one-electron atom[6, 7], although Bohr was unaware that his theory held in *M* rather than *L*. A trickle of papers continue to investigate Bohr's theory, for example Ho,[9] Kastrup,[10] Majumdar and Sharatchandra.[11] The Bohr electron satisfies the free Dirac equation, (2.3.I) with $A^{M\sim} = 0$, and we obtain

$$\phi_1^{M'} = \exp\{i(\eta^\sim x_0^\sim + \mu s)\}, \qquad (3.1.\text{F})$$
$$\phi_2^{M'} = i\{\eta^\sim - \mathbf{i}_s \mu\}\mathbf{M}^{-1}\exp\{i(\eta^\sim x_0^\sim + \mu s)\},$$
$$m^{\sim 2} = \eta^{\sim 2} + \mu^2$$

We have a plane wave solution, $\underline{\Phi}^{M'}(\phi_1^{M'}, \phi_2^{M'})$, and $v = -iv^\sim = -i\mu/\eta^\sim$ is the velocity of the Bohr electron. We comment on why the Bohr electron does not radiate. The circular orbit can be seen as a consequence of the properties of *M* in the same way as a geodesic is a consequence of curvature. Then the electron is in an inertial frame and therefore does not radiate. This depends, however, on considering the electron as a point, so that there are no tidal forces, rather than an extended classical electromagnetic field in *L*.[12]

Although we have an energy eigenstate in the first two equations of (3.1.D) and may operate with a suitable energy operator to obtain the energy eigenvalue for the Bohr interaction,

$$i\frac{\partial \underline{\Phi}^M}{\partial x_0} = iv^\sim \underline{\Phi}^M \qquad (3.1.\text{G})$$

this does not provide a unique answer. However, there is another way of obtaining this energy and that is to consider the Bohr interaction as a whole, the system, rather than the Bohr electron. From the point of view of the system the Bohr electron is travelling with velocity, $-i\tilde{v}$, in $M$. The energy of the system, $i\tilde{v}$, is given in equation (3.1.E) as the sum of the Bohr electron's kinetic energy, $i\tilde{\eta}$, and potential energy, $ieA^{M\sim}$, since the electron is in a potential field. However, from the point of view of the Bohr electron, the system of rest mass, $i\tilde{v}$, is travelling with velocity, $i\tilde{v}$, in $M$. The system as a whole is not in a potential field, this being internal to the bound state, and the energy in the rest frame of the Bohr electron is given simply by the system's kinetic energy which must be equal to the electron's rest mass, $i\tilde{m}$. This implies

$$\tilde{m} = \frac{\tilde{v}}{\sqrt{1+\tilde{v}^2}} \qquad (3.1.\text{H})$$

We may apply de Broglie's relation[8] in $M$ to the Bohr electron described in equation (3.1.F), giving

$$\tilde{\eta} = \frac{\tilde{m}}{\sqrt{1+\tilde{v}^2}} \qquad (3.1.\text{I})$$

Equations (3.1.C), (3.1.E), (3.1.I) and (3.1.H) then give

$$\frac{ief}{R} = \frac{\tilde{m}\tilde{v}^2}{\sqrt{1+\tilde{v}^2}} \qquad (3.1.\text{J})$$

This is the first of Bohr's equations. It is also the same as that proposed by Bohr[6] when his relativistic correction[7] is taken into account.

We may obtain equation (3.1.J) in the absence of a wave function by replacing the electron's mass with the energy-momentum four-vector, but assuming the circular orbit in *L* when $r = R$ is Newtonian. The Newtonian equation of circular motion can then be used.[8] Alternatively, the circular motion can be treated using the geometrical methods of General Relativity, as Bell and Diaz have done.[4] This also leads to equation (3.1.J).

### 3.2  Bohr's second equation

We posit that there is a single value of the wave function, $\underline{\Phi}^M$, at each point in the subspaces of *M* and *L* co-ordinated by $(\theta, r, x_3)$, as is done for the Dirac interaction.[5] Since *s* in the first two equations (3.1.D) follows a circular locus, this implies a condition on the term $i\mu s$. The wave function for the Bohr electron, $\underline{\Phi}^{M'}$ in equation (3.1.F), has the same term. We may therefore apply the condition to the Bohr electron instead. We employ the second of de Broglie's relations for the Bohr electron in *M* to yield

$$\mu = \frac{\tilde{m}\tilde{v}}{\sqrt{1+\tilde{v}^2}} \qquad (3.2.A)$$

After the Bohr electron has completed a circuit, when $s = 2\pi R$, the first and second of equations (3.1.F) enforces

$$R\mu = n \qquad (3.2.B)$$

where *n* is a an integer. Substituting for $\mu$ from equation (3.2.A) into equation (3.2.B) we obtain

$$\frac{m\tilde{\ }\,v\tilde{\ }\,R}{\sqrt{1+v\tilde{\ }^{2}}} = n\frac{h}{2\pi} \qquad (3.2.C)$$

where we have restored $h/2\pi$. This is the second of the Bohr's equations. It is also the same as that proposed by Bohr[6] when his relativistic correction[7] is taken into account.

We note that the condition (3.2.C) means that the descriptor of the spin of the Bohr electron, $\phi_1^M$ or $\phi_2^M$ in the first two equations of (3.1.F), rotates through a circle in $H^C$, the ring of complex quaternions, $n$ times while the vector describing the position of the electron rotates through one circle in both $M$ and $L$. Another way of describing the condition, without explicit reference to the electron's wave function, is to state that the spin descriptor must return to the same orientation for each circuit of the position vector. In other words, the system is cyclic. This route has been taken to satisfy Bohr's second equation in the absence of a wave function by Bell and Diaz.[4]

Equations (3.2.C), (3.1.J) and (3.1.H) give the same energy levels for the Bohr interaction as those proposed by Bohr[6, 7] for the one-electron atom. Exactly the same energy levels are also given by the Dirac interaction applied to the one-electron atom, omitting fine structure.[5] The same energy levels are also given by Sommerfeld's model of the one-electron atom for circular orbits. These occur when the principal and azimuthal quantum numbers are equal. Further details may be found in Eisberg,[8] Stehle[13] and Tomonaga.[14]

## 4. PROPERTIES OF THE BOHR INTERACTION

### 4.1 Energy considerations

The Bohr interaction model also allows us to make statements about the process where the electron moves from one orbit to another in a way the Dirac interaction does not. We discuss what happens in *M* first. We suppose that there is an energy imbalance that means energy has to be transferred into or out of the Bohr interaction, for example, thermally or by the absorption or emission of a photon. We preserve the quantum condition in equation (3.2.C). We have

$$\frac{\tilde{m}\tilde{v}_e R_e}{\sqrt{1+\tilde{v}_e^2}} = n\frac{h}{2\pi} \qquad (4.1.\text{A})$$

where $v_e = -i\tilde{v}_e$ is the velocity of the electron and $R_e$ the appropriate radius. Equations (3.1.C), (3.1.H) and (3.1.I) become

$$\tilde{A} = \frac{if}{R_e}, \qquad \tilde{m} = \frac{\tilde{v}_e}{\sqrt{1+\tilde{v}_e^2}}, \qquad \tilde{\eta}_e = \frac{\tilde{m}}{\sqrt{1+\tilde{v}_e^2}} \qquad (4.1.\text{B})$$

where $ie\tilde{A}$ is the potential energy of the Bohr electron, $i\tilde{\eta}_e$ the kinetic energy of the Bohr electron and $i\tilde{v}_e$ is the total energy for the bound electron. Equation (3.1.E) becomes

$$\Delta\tilde{N} = \tilde{\eta}_e - \tilde{v}_e + e\tilde{A} \qquad (4.1.\text{C})$$

where $i\Delta\tilde{N}$ is the measure of the excess energy.

Substituting from equation (4.1.B) in equation (4.1.C), we find

$$\Delta \tilde{N} = -\frac{\tilde{m}\tilde{v}_e^2}{\sqrt{1+\tilde{v}_e^2}} + \frac{ief}{R_e} \qquad (4.1.D)$$

Substituting from equation (4.1.A) into equation (4.1.D), we obtain

$$\Delta \tilde{N} = \frac{\tilde{m}\tilde{v}_e}{\sqrt{1+\tilde{v}_e^2}} \left( \frac{2i\pi ef}{nh} - \tilde{v}_e \right) \qquad (4.1.E)$$

From the Bohr equations, (3.1.J) and (3.2.C), equation (4.1.E) becomes

$$\Delta \tilde{N} = \frac{\tilde{m}\tilde{v}_e}{\sqrt{1+\tilde{v}_e^2}} \left( \tilde{v} - \tilde{v}_e \right) \qquad (4.1.F)$$

If the energy imbalance, $i\Delta \tilde{N}$, is positive, then equation (4.1.F) shows that $v_e < v$ and equation (4.1.A) shows that $R_e > R$. If this is corrected by the internal energy of the system changing the electron will move to a higher orbit with a larger Bohr radius of $R_e$, with increased $n$. If the energy imbalance, $i\Delta \tilde{N}$, is negative, then equation (4.1.F) shows that $v_e > v$ and equation (4.1.A) shows that $R_e < R$. If this can be and is corrected by the internal energy of the system changing the electron will move to a lower orbit with a smaller Bohr radius of $R_e$, with decreased $n$. Absorption or emission of energy therefore links directly to a change of velocity and Bohr radius which mediate a change in energy level for the Bohr interaction.

We discuss *L*. We could also consider $R_e$ to be the distance of the electron from the nucleus in *L*. If no there is no transfer of energy to or from the environment, then $\Delta \tilde{N} = 0$ and the conservation of energy as implied by equation (4.1.F) forbids the electron to appear anywhere except at a distance from the nucleus equal to the Bohr radius. This occurs because we have taken the nucleus to be a point particle. If instead we augment our model by smearing out the nucleus, for example by including a wave function for it, then this requirement is relaxed. This is true because, roughly speaking, wherever the electron is, there is some non-zero probability that the nucleus is at a distance away equal to the Bohr radius. We discuss one method of smearing out the nucleus below, in sections 4.3 and 4.4, and its representation by a wave function in part (2) of this paper.

### 4.2 Angular momentum eigenvalues

It is possible to take the usual spin operator and lift it into a form applicable to the solution of the versatile Dirac equation (2.1.A) in $L$. The translation of operators is discussed by Bell et al.[1] This provides the usual spin a half eigenvalue and is appropriate if the bispinors, $\phi_1$ and $\phi_2^{\ddagger}$, show half-angular behaviour of the phase under rotation and the analogue under Lorentz transformation. However, we shall find that it is essential that the bispinors, $\phi_1^M$ and $\phi_2^{M\ddagger}$ in the solution of equation (3.1.D), show four-vector behaviour under rotation and Lorentz transformation, as permitted by the versatile photon and Dirac equations. We choose an alternative angular momentum operator measuring the $x_3$-component of the integer orbital angular momentum in $L^{(8)}$

$$\mathbf{m}_{x_3} = ix_1 \frac{\partial}{\partial x_2} - ix_2 \frac{\partial}{\partial x_1} \tag{4.2.A}$$

From equations (2.2.A) and (2.2.C) we obtain

$$\mathbf{m}_{x_3} = i\frac{\partial}{\partial\theta} \qquad (4.2.B)$$

Transforming to *M* using the first of equations (2.2.D) and using equations (3.1.D) and (3.2.A)

$$\mathbf{m}_{x_3}\Phi^B = \frac{\tilde{m}\tilde{v}\tilde{R}}{\sqrt{1+\tilde{v}^2}}\Phi^B \qquad (4.2.C)$$

We see that we have an angular momentum eigenstate and the classical expression for the angular momentum in *L* when $r = R$. From equations (4.2.C) and (3.2.C), the angular momentum of the Bohr interaction, *E*, is *n* in units of $h/2\pi$, where $n \geq 1$.

Although the versatile photon and Dirac equations allow the bispinors to transform like four-vectors, experimentally, the electron's intrinsic spin is found to be a half. Indeed, the angular momentum found for the Dirac interaction is half-integer. Here the orbital angular momentum, *O*, takes an integer value, $|n'| \geq 0$, and the electron spin of a half, *S*, must also be added or subtracted,[5] leading to a total angular momentum, *D*, of $n' \pm 1/2$. In resolution, we suggest that the total angular momentum of the Bohr interaction, *E*, is made up of a contribution from the orbital angular momentum of the Dirac interaction, *O*, the spin a half of the electron, *S*, and a contribution, *B*, from a rotational phase change the electron acquires each orbit because it has been rotated through $2\pi$ radians. We have

$$E = O \pm S \pm B = D \pm B \qquad (4.2.D)$$

We have to add a selection rule to the effect that *E* must not correspond to $n = 0$.

*QED derived from the two-body interaction (1)*

We expand on this. Instead of adding angular momenta we may add phase changes for the electron's bispinors given in the first two of equations (3.1.D) from the point of view of the Dirac interaction. When the electron's position vector has completed one revolution round the nucleus the electron has turned through $2\pi$ radians. The geometrical phase change induced in the electron's bispinors, $\phi_1^M$ and $\phi_2^{M\ddagger}$, in Dirac's spin space for one revolution of the electron is therefore[5]

$$\alpha_1 = \pm\pi \qquad (4.2.E)$$

We consider the ground state. Equations (3.2.A) and (3.2.C) show that the wave length of the electron, $1/\mu$, fits the circumference of the circle traced by the Bohr electron, $2\pi R$, exactly, and the same is true of the electron described in equation (3.1.D). We take half this phase change as due to the electron's intrinsic spin of a half. The phase change induced in the electron's bispinors, $\phi_1^M$ and $\phi_2^{M\ddagger}$, is therefore

$$\alpha_2 = \pm\pi \qquad (4.2.F)$$

where we insist the phase changes in equations (4.2.E) and (4.2.F) have the same sign. The total combined phase change is $\alpha_1 + \alpha_2 = \pm 2\pi$. This is the contribution to the phase of the electron's bispinors, $\phi_1^M$ and $\phi_2^{M\ddagger}$, given in the first and second of equations (3.1.D) for the Bohr interaction. We see from comparing equations (4.2.E) and (4.2.F) that the former is the same as we would expect from another additional spin angular momentum of a half, $B$. We obtain $S + B$ as the total angular momentum which corresponds to a classical orbital angular momentum of one from the point of view of the Bohr interaction. If $O$ is not zero the geometrical phase change in the electron's bispinors, $\phi_1$ and $\phi_2^{\ddagger}$, remains $\alpha_1$ which we continue to associate with an angular momentum, *B*, of a half. Equation (4.2.D) then holds with the total angular momentum of the Bohr interaction, *E*, necessarily being an integer.

We can now understand why the bispinors, $\phi_1^M$ and $\phi_2^{M\ddagger}$, must show four-vector behaviour under rotation and Lorentz transformation. The position vector of the Bohr electron behaves like a four-vector under these transformations. Suppose such a transformation takes place during an orbit. If the bispinors did not have the same four-vector behaviour then they would no longer close the circles which they describe at the same point as the position vector closed the circle it describes. This contradicts equations (3.1.D), (3.2.A) and (3.2.C).

## 4.3 Electron location for a single angular momentum eigenstate

We discuss the shape of the electron's orbit, starting with an eigenstate of $\mathbf{m}_{x_3}$ as given in equations (3.1.D). The wave function does not depend on $x_3$ or *r*. The wave function is also a plane wave although travelling along a curved path, *s*. This means that the electron is equally likely to be anywhere in *M*. Bell et al.[1] derive the Dirac current from which this result follows.

We discuss the location of the electron in *L*. As we described in detail in section 2.3, we perform a Lorentz transformation $Z(x_0, r, \theta, x_3)$ and enter the local frame where the charge density current can be expressed as a charge density. If we have one particle of charge *f* responsible for the charge density, as we have here, we may write $P' = fP$, where $P'$ is the charge density and *P* the probability of finding the particle. We use *P* rather than $P'$. We have

$$\rho^M = \frac{P^M}{\delta V^M} \quad (4.3.A)$$

where $\rho^M$ is the probability density and $P^M$ the probability of finding the electron for *M*. In our calculation in section 3 we assumed the volume element, $\delta V^M$, was constant in *M* which means that all the variables are constant in equation (4.3.A). For *L* at the same point

$$\rho^L = \frac{P^M}{\delta V^L} \quad (4.3.B)$$

where $\rho^L$ is the probability density for *L* and both $\rho^L$ and $\delta V^L$ vary with location in *M*. From equations (4.3.B) and (2.2.O) we obtain

$$\rho^L = \frac{RP^M}{r\delta V^M} \quad (4.3.C)$$

We suppose that at the Bohr radius, when $r = R$, an observer in *L* sees a probability of $P^L = P^M$ and sets $\delta V^L = \delta V^M$ using equation (2.2.O). Since $\delta V^L$ is constant for him $\delta V^L$ remains equal to $\delta V^M$ and he must attribute the variation in $\rho^L$ in equation (4.3.B) to a variation in the probability of finding the electron, $P^L$, and see an apparent probability of

$$P^L = \frac{RP^M}{r} \quad (4.3.D)$$

*r* is the distance of the electron from the $x_3$ axis, which means that the distribution is not spherical.

### 4.4 Electron and nucleus location for superpositions

We consider the shape of the electron probability distribution where the system is not in a single state but in a superposition. The atom appears spherical, for example, during the absorption or emission of a photon. We therefore fix the position of a centre of rotation for the orbits and consider a superposition of orbits of all possible orientations, with each orientation equally likely. We calculate how the average probability of finding the electron in *L*, $\overline{P^L}$, varies under this assumption.

We consider the wave function first, assuming that both the bispinors, $\phi_1^M$ and $\phi_2^{M\ddagger}$, and the mass term vary like a four-vector. The wave function will contain the sum of the wave functions for each individual orientation, with equal amplitudes. Rotations are discussed by Bell et al.[1] Spatial rotations do not alter $\phi_1^M$ which has no spatial component. For the spatial part of $\phi_2^{M\ddagger}$ the sum of the rotations by angle $\xi$ and $\xi + \pi$ radians in every plane will be zero and therefore only the scalar part of $\phi_2^{M\ddagger}$, $\phi_2$ survives, again unchanged. Thus the wave function for a superposition is also a plane wave. In particular we do not get quantum interference between the various states if we consider the expectation value for an operator such as position. Our discussion in the last section remains valid for a superposition for some fixed *r*.

We shall place the centre of rotation for the orbits at $x_r = 0$ which we call point *G*. Let the electron be at a point *H* a distance *a* from *G*. Let the intersection of the $x_3$ axis with the sphere centred on *G* of radius *a* be $H'$. Let the foot of the perpendicular from *H* onto the line $GH'$ be *I*. *I* is the location of the nucleus if the Bohr interaction is in the state described by equation (3.1.D). We now turn to the superposition of orbits. We wish to calculate the average length of *HI*, in other words the average distance from the point *H* to the line through *GH'*, when we allow *H'* to vary over the surface of the sphere. The distance from *H* to *GH'* is the same as the distance from *H'* to the line through *GH*. Instead of allowing *H'* to vary we fix *H'* and allow the location of *H* to vary over the surface of the sphere. Let angle $HGH' = \vartheta$. We may calculate the surface area of the sphere, *K*, by finding the integral of the surface of revolution, $2\pi a^2 \sin\vartheta\, \delta\vartheta$, of the line element $a\delta\vartheta$ at *G*

$$K = \int_0^\pi 2\pi a^2 \sin\vartheta \, d\vartheta = 4\pi a^2 \qquad (4.4.A)$$

where $r = a\sin\vartheta$ is the distance from *H* to the line *GH'*. The average distance, $\bar{r}$, of *H* from the line and average inverse distance, $\overline{1/r}$, is then given by

$$\bar{r} = \frac{\int_0^\pi 2\pi a^3 \sin^2\vartheta \, d\vartheta}{4\pi a^2}, \qquad \overline{1/r} = \frac{\int_0^\pi 2\pi a \, d\vartheta}{4\pi a^2} \qquad (4.4.B)$$

Performing these calculations, we find that

$$\bar{r} = \frac{\pi a}{4}, \qquad \overline{1/r} = \frac{\pi}{2a} \qquad (4.4.C)$$

We see that both $\bar{r}$ and the reciprocal of $\overline{1/r}$, are proportional to *a*, the distance of the electron from *G*.

Substituting $\overline{1/r}$ for $1/r$ from the second of equations (4.4.C) into equation (4.3.D) we obtain

$$\overline{P^L} = \frac{\pi R P^M}{2a} \qquad (4.4.\text{D})$$

The distribution is now spherically symmetrical and the electron is now more likely to be found near *G*. Our averaging procedure above smeared the nucleus over the points inside the sphere of radius *a* centred on *G* and this results in the average position of the nucleus being at point *G*. We used equation (3.1.B) for the strength of the potential in *L* as our starting point for deriving the Bohr interaction. Let us assume equation (3.1.B) for single states as given by equation (3.1.D). If we substitute $\overline{1/r}$ for $1/r$ in equation (3.1.B) and then for $\overline{1/r}$ from equation (4.4.C) we obtain

$$\overline{A}\tilde{\phantom{a}} = \frac{if\pi}{2a} \qquad (4.4.\text{E})$$

where $i\overline{A}\tilde{\phantom{a}}$ is the average potential and we see we have turned the potential in (3.1.B) into an inverse distance from the nucleus law for superpositions. This is the same as the assumption made for the Dirac interaction, apart from a constant of proportionality.

### 4.5 Angular momentum for superpositions

As we saw in our discussion of the wave function for superpositions, and Bell et al. remarked,[1] rotation by $\pi$ does not change a quantum-mechanical spin up state into a spin down state when the bispinors, $\phi_1^M$ and $\phi_2^{M\ddagger}$, and the mass term vary like a four-vector. This means that a superposition continues to be an eigenstate of $\mathbf{m}_{x_3}$ for the Bohr interaction. The transition between a quantum mechanical spin up and spin down state is only effected when the mapping between the spin and *M* is altered.[1] In our current case the mapping takes place in equation (3.2.B). We chose a positive value of *n* but a negative value is equally permissible. Had we chosen a negative value of *n* equations (4.2.C) and (3.2.C) would have led to a negative value for the angular momentum. We have to choose a method of deciding which mapping applies.

The phase change due to the rotation of an electron is classical in the sense it depends upon a classical curved path or rotation on the spot. Classically, for each orientation of the plane of the electron's orbit a clockwise and an anticlockwise rotation is equally likely. The electron is therefore on average at rest. We suppose that for each orbit the clockwise rotation is mapped to spin up, while the anticlockwise rotation is mapped to spin down and that the electron alternates between the two. The phase change in the bispinors, $\phi_1^M$ and $\phi_2^{M\ddagger}$, due to the rotation of the electron position vector is opposite for the two possibilities, leading to an average contribution to the angular momentum of zero. The total angular momentum associated with each plane in the superposition becomes *D* as in the Dirac interaction.

This brings us from the cylindrical Bohr interaction model in *M* back full circle to the spherical Dirac interaction model in *L*.

## 5. CONCLUSION

We compare the Bohr interaction in *M*-space to the usual model, the Dirac interaction in *L*. We see that the Bohr interaction produces a state with energy and angular momentum eigenvalues identical to a subset of those found with the Dirac interaction. The Bohr interaction also has similarly local properties for the probability density in *L*. The Bohr interaction even predicts the applicability of the Dirac interaction in *L* with its inverse distance spherically-symmetrical potential. These factors lead us to conclude that both models are physically viable and that there is nothing preventing a Dirac interaction from becoming a Bohr interaction or vice versa, except possibly the symmetry of the boundary conditions.

This leaves the fine structure predicted by the Dirac interaction out of account. Bell et al.[3] and Bell and Diaz[4] build on the results presented here and consider a similar model for an attractive two-body interaction but this time with a circular orbit with two circular components. Both components obey the Bohr equations. This model then incorporates the Bohr interaction as described here and predicts the fine structure spectrum in addition accounting for all the possible eigenvalues of the bound system.

**ACKNOWLEDGEMENTS**

One of us (Bell) would like to acknowledge the assistance of E.A.E. Bell.